\documentclass[conference,compsoc]{IEEEtran}

\usepackage{balance}
\def\Goal{%
to raise software developers' awareness of the security implications when selecting code snippets%
by empirically comparing the vulnerabilities of ChatGPT and StackOverflow%
}

\def\RQOne{What vulnerabilities differences are there between ChatGPT and \gls{SO} code snippets?}
\def\RQTwo{What types of vulnerabilities in terms of \gls{CWE} types are present for ChatGPT-generated code versus \gls{SO}-answered?}

\ifCLASSOPTIONcompsoc
  \usepackage[nocompress]{cite}
\else
  \usepackage{cite}
\fi

\usepackage[export]{adjustbox}
\usepackage{enumitem}
\usepackage{hyperref}
\usepackage{standalone}
\usepackage{tikz}
\usetikzlibrary{positioning}
\usepackage{tabularx}
\usepackage{listings}
\usepackage{booktabs}
\usepackage{xpatch}
\usepackage{xcolor,colortbl}
\usepackage{listings}
\usepackage{realboxes}
\usepackage{upquote}

\usepackage[most]{tcolorbox}

\tcbset{colback=black!5!white,colframe=black}

\newcommand{\winnertable}[1]{\cellcolor{gray!25}{#1}}
\definecolor{mygray}{rgb}{0.93,0.93,0.93}

\newcommand{\codebox}[1]{\colorbox{mygray}{\lstinline|#1|}}

\usepackage[acronym]{glossaries}

\newacronym{AI}{AI}{Artificial Intelligence}
\newacronym{LLM}{LLMs}{Large Language Models}
\newacronym{ML}{ML}{Machine Learning}
\newacronym{SO}{SO}{StackOverflow}
\newacronym{API}{API}{StackOverflow}
\newacronym{GQM}{GQM}{Goal-Question-Metric}
\newacronym{CWE}{CWE}{Common Weakness Enumeration}

\def\gptversion{\codebox{gpt-3.5-turbo-0613}}

\def\finalprompt{\textit{For all the following questions, generate a compilable code snippet in Java. [BODY].}}

\def\codeqlfile{\codebox{java-security-and-quality.qls}}
\def\codeqlversion{\codebox{v2.16.1}}

\newcommand{\NumberInitialDatasetAnswers}{1,429}
\newcommand{\YearsDatasetPosts}{2008-2017}

\newcommand{\NumberQuestionsInitialPurposeful}{1,216}
\newcommand{\NumberAnswersInitialPurposeful}{1,377}
\newcommand{\NumberSnippetsInitialPurposeful}{3,739}
\newcommand{\NumberAnswersInitialPurposefulError}{52}

\newcommand{\NumberQuestionsFinalPurposeful}{189}
\newcommand{\NumberAnswersFinalPurposeful}{204}
\newcommand{\NumberSnippetsFinalPurposeful}{216}

\newcommand{\NumberSOGPTPairs}{237}

\newcommand{\NumberFinalQuestions}{87}
\newcommand{\NumberFinalAnswers}{90}
\newcommand{\NumberFinalSnippets}{108}

\newcommand{\NumberUniqueVulOverall}{274}

\newcommand{\NumberGPTVulQuestion}{77}

\newcommand{\NumberGPTVulAnswer}{79}

\newcommand{\NumberGPTVulSnippet}{87}

\newcommand{\NumberGPTVul}{248}
\newcommand{\NumberGPTVulAvg}{3.04}
\newcommand{\NumberGPTVulMedian}{2}

\newcommand{\NumberGPTVulUnique}{158}

\newcommand{\NumberSOVulQuestion}{75}

\newcommand{\NumberSOVulAnswer}{77}

\newcommand{\NumberSOVulSnippet}{83}

\newcommand{\NumberSOVul}{302}
\newcommand{\NumberSOVulAvg}{3.62}
\newcommand{\NumberSOVulMedian}{2}
\newcommand{\NumberSOVulMax}{24}

\newcommand{\NumberSOVulUnique}{183}

\newcommand{\NumberPVulQues}{p = 0.87}
\newcommand{\NumberPVulAnswer}{p = 0.87}
\newcommand{\NumberPVulSnippets}{p = 0.76}
\newcommand{\NumberPVul}{p = 0.02}
\newcommand{\NumberPVulUnique}{p = 0.18}
\newcommand{\NumberPVulSnippetAvg}{p = 0.25}

\newcommand{\NumberDifferenceBoth}{20\%}

\newcommand{\NumberOverlapPerQuestions}{81\%}

\newcommand{\NumberOverlapPerAnswers}{79\%}

\newcommand{\NumberOverlapPerSnippets}{79\%}

\newcommand{\NumberOverlapPerVulnerabilitiesUnique}{25\%}

\newcommand{\NumberOverlapPerVulnerabilitiesUniqueDiff}{54\%}

\newcommand{\NumberTotalCWEType}{25}
\newcommand{\NumberGPTCWEType}{19}
\newcommand{\NumberSOCWEType}{22}

\newcommand{\NumberGPTCWETypeBetter}{15}
\newcommand{\NumberGPTCWETypeNeither}{2}
\newcommand{\NumberSOCWETypeBetter}{8}

\newcommand{\NumberDifferenceTypeBoth}{14.63\%}

\begin{document}
\title{Just another copy and paste? Comparing the security vulnerabilities of ChatGPT generated code and StackOverflow answers}
\author{\IEEEauthorblockN{Sivana Hamer, Marcelo d'Amorim, Laurie Williams}
\IEEEauthorblockA{
Department of Computer Science, North Carolina State University, Raleigh, North Carolina\\
Email: sahamer@ncsu.edu, mdamori@ncsu.edu, lawilli3@ncsu.edu}
}

\maketitle
\thispagestyle{plain}
\pagestyle{plain}

\begin{abstract}
Sonatype's 2023 report found that 97\% of developers and security leads integrate generative \gls{AI}, particularly \gls{LLM}, into their development process.  
Concerns about the security implications of this trend
have been raised.
Developers are now weighing the benefits and risks of \gls{LLM} against other relied-upon information sources, such as \gls{SO}, requiring empirical data to inform their choice.
In this work, our goal is \Goal.
To achieve this, we used an existing Java dataset from \gls{SO} with security-related questions and answers. 
Then, we asked ChatGPT the same \gls{SO} questions, gathering the generated code for comparison.
After curating the dataset, we analyzed the number and types of \gls{CWE} vulnerabilities of \NumberFinalSnippets{} snippets from each platform using CodeQL.
ChatGPT-generated code contained \NumberGPTVul{} vulnerabilities compared to the \NumberSOVul{} vulnerabilities found in \gls{SO} snippets, producing \NumberDifferenceBoth{} fewer vulnerabilities with a statistically significant difference.
Additionally, ChatGPT generated \NumberGPTCWEType{} types of \gls{CWE}, fewer than the \NumberSOCWEType{} found in \gls{SO}.
Our findings suggest developers are under-educated on insecure code propagation from both platforms, as we found \NumberUniqueVulOverall{} unique vulnerabilities and \NumberTotalCWEType{} types of \gls{CWE}.
Any code copied and pasted, created by \gls{AI} or humans, cannot be trusted blindly, requiring good software engineering practices to reduce risk. 
Future work can help minimize insecure code propagation from any platform.
\end{abstract}

\glsresetall
\section{Introduction}\label{SEC:Introduction}

\gls{AI} as assistant tools have become commonplace within software development.
According to Sonatype's 2023 State of the Supply Chain report~\cite{SonarType2023Annual},
$97\%$ of developers and security leads integrate generative \gls{AI} into their software engineering processes.
In particular, the report found that \gls{LLM} are commonly utilized by developers, with ChatGPT as the most used tool.
Moreover, GitHub's State of Open Source Software report for 2023~\cite{github2023State} revealed that almost a third of open-source projects with stars have a maintainer using GitHub Copilot.

As \gls{LLM} are now widely adopted within software engineering, practitioners have raised concerns regarding the security of the tool~\cite{enck2023S3C2,dunlap2023S3C2}.
Code generated by \gls{LLM} can contain vulnerabilities~\cite{pearce2022Asleep, liu2023Need} and security-related code smells~\cite{siddiq2022Empirical}. 
Research has found vulnerable code generated by \gls{LLM} in GitHub~\cite{fu2023Security}. 
Consequently, \gls{AI}-generated code may affect the software supply chain.
Sonatype's 2023 report on Generative AI in software development~\cite{SonarType2023Risks} highlighted that security is a primary concern among respondents. 
Alarmingly, despite security concerns, $74\%$ of respondents felt pressured into using generative \gls{AI}.

\begin{figure}[tb]
   \centering
    {
        \includegraphics[width=\linewidth]{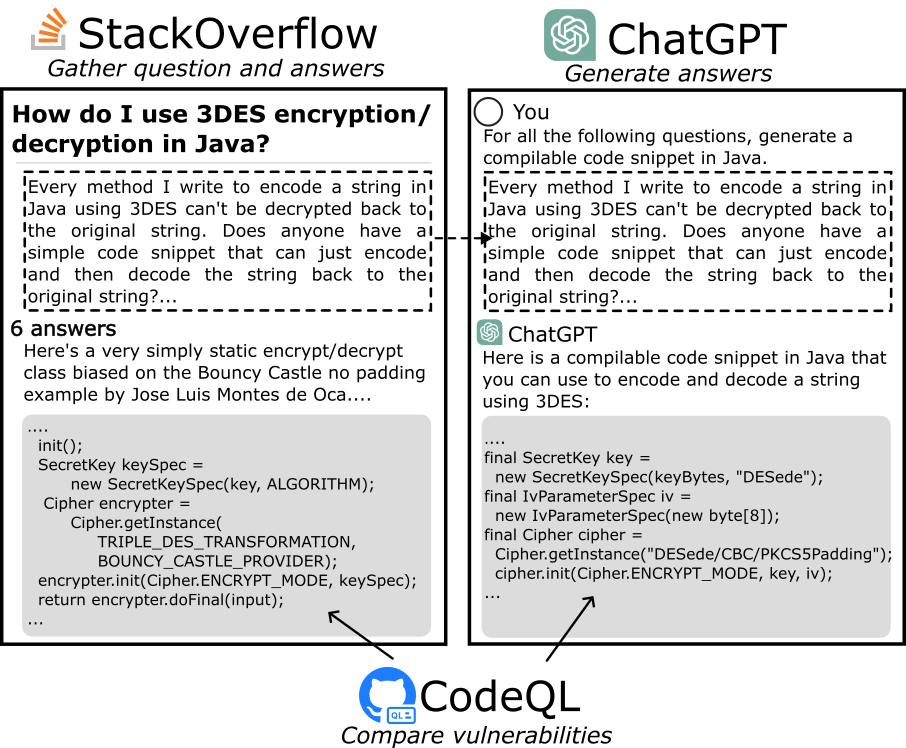}
    }
    \vspace{-3ex}
    \caption{Example of how we compared ChatGPT and StackOverflow}\label{FIG:ExampleComparison}
    \vspace{-2ex}
\end{figure}

Developers are weighing the benefits and risks of using \gls{LLM} by contrasting them with other information sources, notably online forums like \gls{SO}, commonly used during development~\cite{kabir2023Answers, xu2023We}.
Consequently, a developer may want to know how the security of \gls{LLM}-generated code compares with other web-based information sources.
The choice of development source influences the security and functionality of the resulting code~\cite{acar2016You}.
As current web-based information sources contain insecure code~\cite{fischer2017Stack, chen2019How, rahman2019Snakes, zhang2022Study}, developers must know which information source is less insecure. 
Hence, choosing a more secure source can propagate fewer security issues to developers and guide policy decisions about using \gls{LLM}.
Empirical data is needed to provide insights about the differences in code vulnerabilities between \gls{LLM} and other information sources for informed decision-making.

The goal of this work is \textit{\Goal}.
To achieve this goal, the following research questions were answered:

\begin{enumerate}[topsep=1pt, leftmargin=*, itemsep=2pt, label=\textbf{RQ\arabic*}:, ref=\textbf{RQ\arabic*}]
\item \RQOne
\item \RQTwo
\end{enumerate}

To answer these questions, we conducted an experimental study comparing two widely adopted web-based information sources used in software development: ChatGPT~\cite{openAI2024ChatGPT} and \gls{SO}~\cite{so2024SO}.
Fig.~\ref{FIG:ExampleComparison} provides an example of how we compared the ChatGPT and \gls{SO}.
In this study, we used \NumberFinalSnippets{} Java security-related code snippets from \gls{SO} taken from Chen et al.~\cite{chen2019How}.
We queried ChatGPT, reusing the \gls{SO} question as our prompt, to gather \gls{AI}-generated code snippets for comparison.
We then detected the vulnerabilities in the code with CodeQL~\cite{github2024CodeQL}, a static analysis tool developed by GitHub that provides \gls{CWE}.

We found that ChatGPT-generated code contains fewer vulnerabilities and types of vulnerabilities compared to \gls{SO}.
Still, both platforms can be sources of insecure code propagation as we found \NumberUniqueVulOverall{} unique vulnerabilities and \NumberTotalCWEType{} types of \gls{CWE}. 
Additionally, the vulnerabilities found in the ChatGPT-generated code and \gls{SO} overlapped only in \NumberOverlapPerVulnerabilitiesUnique{} of the vulnerabilities.
The contributions of our work are:

\begin{itemize}
    \item A comparison of the vulnerabilities found in \NumberFinalSnippets{} of ChatGPT-generated and \gls{SO}-answered Java security-related code snippets.
    \item An enhanced dataset of \NumberFinalSnippets{} ChatGPT snippets  associated with \NumberFinalQuestions{} questions and \NumberFinalAnswers{} answers from security-related \gls{SO} questions~\footnote{Our dataset: \url{https://zenodo.org/records/10806611}}.
    \item A list of \NumberTotalCWEType{} different vulnerabilities found in ChatGPT or \gls{SO} snippets with their respective \gls{CWE}s.
\end{itemize}

The remainder of this work is divided as follows. 
Section~\ref{SEC:Methodology} details the methodology.
Section~\ref{SEC:Results} describes our results and Section~\ref{SEC:Discussion} discusses our findings.
Section~\ref{SEC:Limitations} delineates the limitations of the study.
Section~\ref{SEC:Related} presents the related work. 
Finally, Section~\ref{SEC:Conclusions} concludes the work.
\section{Methodology}\label{SEC:Methodology}

We conducted an experimental study to compare both information sources in five steps.
First, we selected the platforms under study, ChatGPT and \gls{SO} (Step 1).
We then selected security-related questions and answers from \gls{SO} (Step 2) and collected the code snippets from the answers (Step 3).
We prompted ChatGPT with the \gls{SO} questions to generate code (Step 4) and gathered the generated snippets (Step 3).
We then compare the gathered vulnerabilities of \gls{SO} and ChatGPT using CodeQL (Step 5).
In the following subsections, we detail each step.

\subsection{Step 1: Platform selection}\label{SUBSEC:MET-CASE}

We choose the two platforms to compare the code in \gls{LLM} and web-based information sources.
For our LLM we chose ChatGPT, developed by OpenAI, for two main reasons.
First, in the 2023 report on Generative AI in software development, Sonatype found that ChatGPT was the most used tool by $86\%$ of the respondents~\cite{SonarType2023Risks}.
Second, \gls{LLM} can be interacted with as a conversational agent in a similar way that a developer may ask a question and receive an answer in \gls{SO}.
The tools are thus directly comparable.
For our traditional web-based information source we selected \gls{SO}, an online question-and-answer forum for programmers, for the following reasons.
\gls{SO} remains widely utilized and frequented by its users.
Based on data from February 2024, the site hosts 22 million users with 2,700 questions per day~\cite{stackexchange2024}.
$92.5\%$ of users visit the site at least weekly or a few times a month~\cite{so2023Survey}.
Additionally, prior work about insecure code propagation has investigated the \gls{SO} site~\cite{fischer2017Stack, chen2019How, rahman2019Snakes, zhang2022Study, acar2016You}.
By choosing \gls{SO}, we directly build upon and extend prior knowledge on the security of online information sources.

\subsection{Step 2: Question and answer selection}\label{SUBSEC:MET-QA}

We collect a dataset of questions with answers to sample \gls{SO} answers and generate code with \gls{LLM}.
We intentionally sampled the questions and answers to serve the purpose of our study through a purposeful sample~\cite{baltes2022Sampling}.
We used a previously curated Java code snippet dataset by Chen et al.~\cite{chen2019How}. 
The dataset provided \NumberInitialDatasetAnswers{} \gls{SO} answers with secure or insecure code snippets for security-related questions from \YearsDatasetPosts{}.
The dataset was chosen as it was previously manually curated and is publicly available.
Additionally, a dataset with security-related questions was selected as not all questions in the wild have software vulnerabilities.
Our sampling approach can thus serve as an upper bound for vulnerabilities found in \gls{SO} posts.
At the same time, Java is consistently a top programming language for developers in open source software~\cite{github2023State}.
After gathering the answers identifiers from the datasets, when analyzing the data we noticed that some snippets were partially stored.
For example, imports were missing from the snippet in \gls{SO}.
At the same time, the dataset did not store the text of the associated questions and answers.
As we needed the complete snippets, we mined \gls{SO} to gather the answers through the StackExchange API~\cite{api2024StackExchange}.
Specifically, we collected the title and body of the questions with the answers.
In total, \NumberQuestionsInitialPurposeful{} questions and \NumberAnswersInitialPurposeful{} answers were mined.
We could not gather \NumberAnswersInitialPurposefulError{} answers indicated in the dataset with their respective questions as they were no longer available in \gls{SO}.

\subsection{Step 3: Code snippet filtration}\label{SUBSEC:MET-SNIPPET}

We curate the code snippets to analyze the answers' vulnerabilities.
Hence, we collected the associated code snippets for each of the \NumberAnswersInitialPurposeful{} answers. 
To achieve this in an automated manner, we utilized the \codebox{Beautiful Soup 4} Python library~\cite{beautiful2024Soup} to parse the HTML of the answers to find the code blocks.
We further checked the found code blocks to verify they were Java snippets using the \codebox{javalang} Python library~\cite{javalang2024Javalang}.
We gathered in total \NumberSnippetsInitialPurposeful{} code snippets.

We further curated our data to verify that the snippets could compile further analysis (Step 5).
To find code snippets that were not single lines of code, as one-liners are not compilable in Java, we gathered snippets stored in code blocks.
Based on our analysis, the criteria were verified if the code block's parent HTML tag was of type \codebox{<pre>}.
After applying the criteria, 2,432 snippets remained.
As code snippets may have multiple Java classes in the same snippet or none at all, we then filtered that snippets had only one class name.
We verified the criteria using the \codebox{javalang} Python library~\cite{javalang2024Javalang}.
After the step, 526 snippets remained.

Finally, we compiled each snippet to satisfy the vulnerability detection tool requirements.
We did not modify any code snippets to avoid introducing bias into the code.
Hence, we verified and tried to fix compilation errors due to missing package libraries for all close snippets with a missing package error.
In our initial compilation of the snippets, we gathered all errors due to missing packages through the regular expression \codebox{error: package [a-zA-z\.]+ does not exist}.
We manually searched and downloaded all missing packages through Maven the \codebox{.jar} files, searching all packages.
Then, we recompiled all the snippets.
We were unsuccessful for 62 packages missing in 86 code snippets, which may have also contained other compilation errors.
In total, \NumberSnippetsFinalPurposeful{} snippets remained, corresponding to \NumberQuestionsFinalPurposeful{} questions and \NumberAnswersFinalPurposeful{} answers. 

\subsection{Step 4: ChatGPT answers generation}\label{SUBSEC:MET-GEN}

We generate answers from ChatGPT to compare \gls{LLM} generated code. 
For each of the \NumberSnippetsFinalPurposeful{} remaining snippets, we asked the ChatGPT model the following prompt: 
\finalprompt

We selected a prompt that leveraged the \gls{SO} post as we did not want to introduce additional bias to the model.
Still, based on our tests, we were required to detail which programming language the question was for, which in our case was Java. 
Additionally, to verify that the code given was a complete program and not a code snippet, we specified that the code must be compilable due to the limitations of our vulnerability detection tool for Java. 
Finally, we prompted the model with the complete body of the \gls{SO} question, though more expensive, was more comparable to how users asked questions in \gls{SO}.

We thus prompted ChatGPT (\gptversion) through the OpenAI REST API in Python~\cite{openai2024Python}
and stored each reply.
Combining the answers with the questions generated a total of \NumberSOGPTPairs{} pairs of \gls{SO} snippets. 
We then followed the process outlined in Step 3 with some minor differences.
First, as the API returned markdown, we processed the result with the regular expression \codebox{```[^`]*```} finding code blocks based on whether the text was enclosed by a code snippet block (\codebox{```}).
Second, as we only selected code block snippets, we did not need to filter if they were code blocks.
Additionally, we downloaded missing libraries for the ChatGPT snippets with the same procedure.
We were unsuccessful for 29 imports for 11 code snippets that could have other compilation errors.
After filtering \NumberFinalSnippets{} snippets pairs related to \gls{SO} \NumberFinalQuestions{} questions and \NumberFinalAnswers{} answers remained.

\subsection{Step 5: Vulnerability detection}\label{SUBSEC:MET-VUL-DET}

We utilized a static analysis tool to determine the security vulnerabilities of each code snippet from ChatGPT and \gls{SO}.
These tools are one way to detect vulnerabilities and have been used in industry~\cite{elder2022really}. 
Specifically, we utilized CodeQL~\cite{github2024CodeQL}, a semantic code analysis engine developed by GitHub.
CodeQL has been used to evaluate the code generated by \gls{LLM} in prior research~\cite{pearce2022Asleep, fu2023Security}.
At the same time, CodeQL is one of the better-performing static analysis tools for Java~\cite{li2023Comparison}.
CodeQL detects vulnerabilities by querying the code on a database with vulnerability variants.
We utilized version \codeqlversion{} and \codeqlfile{} test suite that contains queries for security and quality of Java code.
We analyzed 234 queries for each snippet.

To determine the significance of the differences during our analysis, we utilized standard statistical analysis tests in R~\cite{kloke2014Nonparametric}.
We mapped the \gls{CWE} we found with MITRE's 2023 Top-25 Most Dangerous Software Weaknesses list~\cite{mitre2023Top}.

\section{Results}\label{SEC:Results}

\subsection{Vulnerabilities}

\begin{table}[tb]
    \centering
    \caption{Summary of vulnerabilities in ChatGPT and StackOverflow (SO).
     The platform with fewer vulnerabilities is highlighted.}
    \begin{tabular}{lrrr}
        \toprule
         & \textbf{GPT} & \textbf{SO} & \textbf{Overlap} \\
        \midrule
        Questions with vulnerabilities & \NumberGPTVulQuestion & \winnertable{\NumberSOVulQuestion} & \NumberOverlapPerQuestions \\
        Answers with vulnerabilities & \NumberGPTVulAnswer & \winnertable{\NumberSOVulAnswer} & \NumberOverlapPerAnswers  \\
        Code snippets with vulnerabilities & \NumberGPTVulSnippet & \winnertable{\NumberSOVulSnippet} & \NumberOverlapPerSnippets \\
        Vulnerabilities in snippets* & \winnertable{\NumberGPTVul} & \NumberSOVul & - \\
        Unique vulnerabilities in snippets & \winnertable{\NumberGPTVulUnique} & \NumberSOVulUnique & \NumberOverlapPerVulnerabilitiesUnique\\
        \bottomrule
    \end{tabular}\\
    \vspace{.5ex}
    {\scriptsize Chi-squared test significance: `***' $p <0.001$, `**' $p <0.01$, `*' $p <0.05$}
    \vspace{-3ex}
    \label{TAB:SummaryVul}
\end{table}

To understand the differences between platforms, we start by analyzing the vulnerabilities detected by CodeQL.
Table~\ref{TAB:SummaryVul} shows a summary of the results. 
We found vulnerabilities in \NumberGPTVulQuestion{} questions for ChatGPT. 
On the other hand, the number of questions with vulnerabilities 
for \gls{SO} was \NumberSOVulQuestion{}.
Hence, \gls{SO} had fewer questions with vulnerabilities, yet the differences were insignificant when we performed a Chi-squared test (\NumberPVulQues{}). 
Additionally, the overlap in questions with vulnerabilities was \NumberOverlapPerQuestions{}.
Similarly, the number of answers with vulnerabilities was \NumberGPTVulAnswer{} for ChatGPT and \NumberSOVulAnswer{} for \gls{SO}, with an overlap of \NumberOverlapPerAnswers{}.
\gls{SO} had fewer answers with vulnerabilities, yet the differences were not statistically significant (\NumberPVulAnswer{}).
Likewise, ChatGPT generated \NumberGPTVulSnippet{} snippets with vulnerabilities, higher than the \NumberSOVulSnippet{} found in \gls{SO} with an overlap of \NumberOverlapPerSnippets{}.
The differences were not statistically significant (\NumberPVulSnippets{}).

We found \NumberGPTVul{} vulnerabilities across all ChatGPT code snippets.
Meanwhile, we found \NumberSOVul{} vulnerabilities in the \gls{SO} code snippets. 
Hence, the number of vulnerabilities is \NumberDifferenceBoth{} fewer in ChatGPT snippets than \gls{SO}. 
We performed a Chi-squared test to determine if the differences in the number of vulnerabilities of the code snippets by platform were statistically significant. 
We found a statistically significant difference (\NumberPVul).

\begin{figure}[tb]
   \centering
    {
        \includegraphics[width=\linewidth]{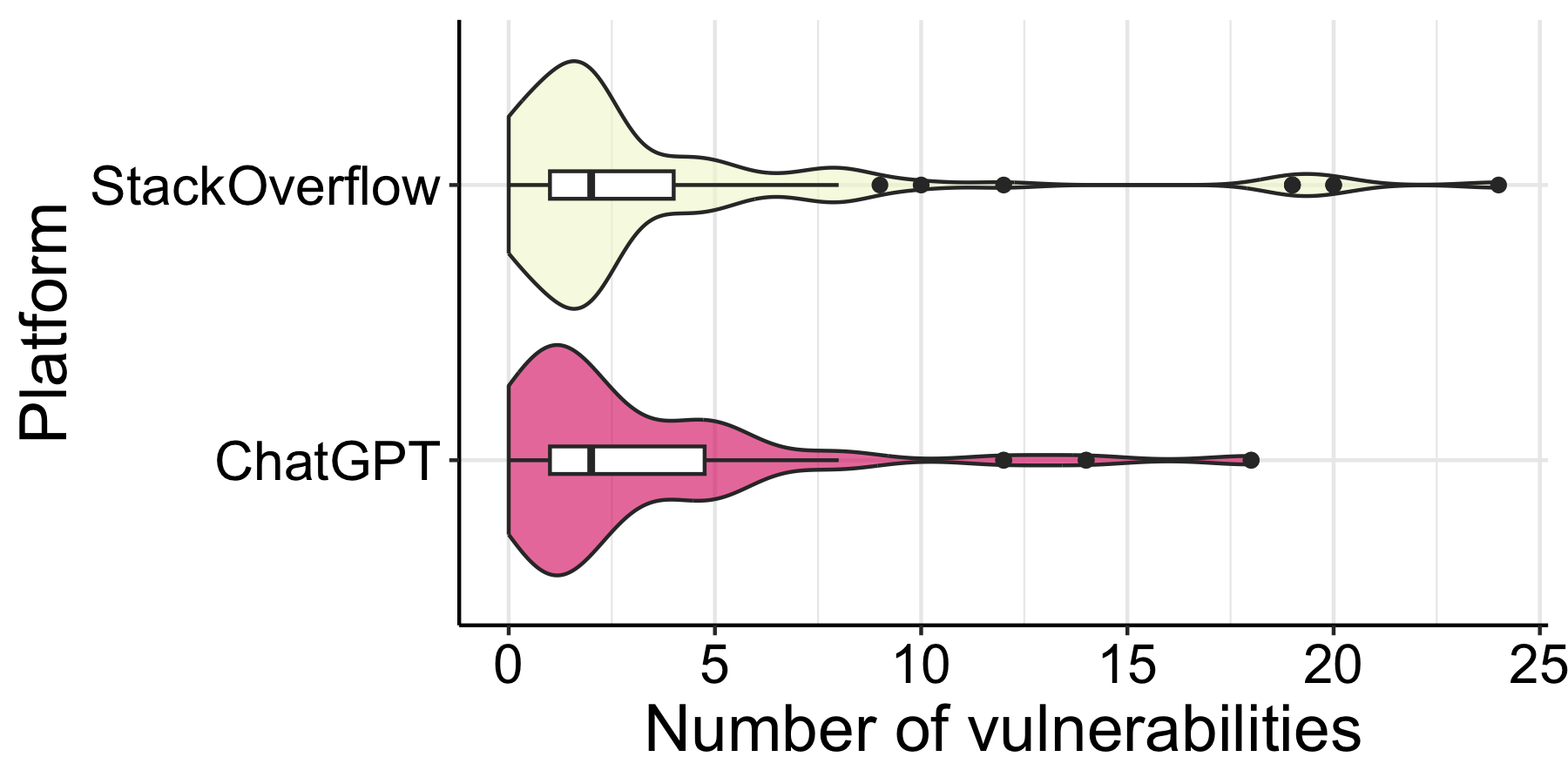}
    }
    \vspace{-3ex}
    \caption{Number of vulnerabilities in the code snippets in each platform.}\label{FIG:BoxVulnerability}
\end{figure}

Fig.~\ref{FIG:BoxVulnerability} shows a violin chart with a box plot of the number of vulnerabilities of the snippets for each platform.
The average and median number of vulnerabilities per snippet for ChatGPT were \NumberGPTVulAvg{} and \NumberGPTVulMedian{}, respectively. 
The average number of vulnerabilities for \gls{SO} code snippets was \NumberSOVulAvg{}, while \NumberSOVulMedian{} was the median.
The maximum vulnerabilities found in a snippet was \NumberSOVulMax{}, created in \gls{SO}. 
To compare the number of vulnerabilities in each code snippet produced in \gls{SO} versus ChatGPT, we utilized a paired t-test.
We did not find a statistically significant difference in the number of vulnerabilities in each code snippet (\NumberPVulSnippetAvg{}). 
Therefore, the differences were not statistically significant despite ChatGPT producing on average code snippets with fewer vulnerabilities than \gls{SO}.

We gathered the unique vulnerabilities in each snippet to compare the overlap of vulnerabilities, as vulnerabilities could be present in different lines.
The number of unique vulnerabilities in the code snippets is \NumberGPTVulUnique{} for ChatGPT and \NumberSOVulUnique{} for \gls{SO}. 
We found no statistically significant difference with the unique snippet vulnerabilities for the platforms using a Chi-squared test (\NumberPVulUnique).
The vulnerabilities generated by ChatGPT and present in \gls{SO} code snippets are only the same in \NumberOverlapPerVulnerabilitiesUnique{} of snippets.
Noticeably, there is a difference of at least \NumberOverlapPerVulnerabilitiesUniqueDiff{} between the overlap of unique vulnerabilities in snippets and the other overlaps. 

\subsection{Vulnerabilities CWEs}

\begin{table}[tb]
    \centering
    \caption{The types of \gls{CWE} from the information sources. The platform with fewer vulnerabilities is highlighted. The delta ($\Delta$) represents the difference between platforms.}
    \begin{tabular}{lrrrr}
        \toprule
        \textbf{CWE-ID} & \textbf{\# GPT} & \textbf{\# SO} & \textbf{$\Delta$} & \textbf{Top 25} \\
        \midrule
        CWE-078 & 2 & \winnertable{0} & 2 & 5 \\
        CWE-088 & 2 & \winnertable{0} & 2 & - \\
        CWE-248 & \winnertable{4} & 10 & 6 & - \\
        CWE-295 & 13 & \winnertable{12} & 1 & - \\
        CWE-297 & 4 & 4 & 0 & - \\
        CWE-326 & \winnertable{4} & 7 & 3 & - \\
        CWE-327 & \winnertable{95} & 122 & 27 & - \\
        CWE-328 & \winnertable{95} & 122 & 27 & - \\
        CWE-329 & \winnertable{7} & 8 & 1 & - \\
        CWE-330 & \winnertable{0} & 3 & 3 & - \\
        CWE-335*** & \winnertable{0} & 16 & 16 & - \\
        CWE-338 & \winnertable{0} & 3 & 3 & - \\
        CWE-391 & \winnertable{14} & 18 & 4 & - \\
        CWE-404 & 37 & \winnertable{29} & 8 & - \\
        CWE-476 & \winnertable{1} & 6 & 5 & 12 \\
        CWE-477 & 19 & \winnertable{15} & 4 & - \\
        CWE-561 & 13 & 13 & 0 & - \\
        CWE-570* & \winnertable{0} & 4 & 4 & - \\
        CWE-571* & \winnertable{0} & 4 & 4 & - \\
        CWE-581 & \winnertable{0} & 2 & 2 & - \\
        CWE-772 & 37 & \winnertable{29} & 8 & - \\
        CWE-780 & \winnertable{3} & 6 & 3 & - \\
        CWE-798* & 36 & \winnertable{20} & 16 & 18 \\
        CWE-835** & 8 & \winnertable{0} & 8 & - \\
        CWE-1204 & \winnertable{7} & 8 & 1 & - \\
    \bottomrule
    \end{tabular}\\
    \vspace{.5ex}
    {\scriptsize Chi-squared test significance: `***' $p <0.001$, `**' $p <0.01$, `*' $p <0.05$}
    \vspace{-3ex}
    \label{TAB:CWEs}
\end{table}

We gathered the vulnerabilities associated with \gls{CWE}s to understand how the security issue types varied between platforms. 
The \gls{CWE}s we found are shown in Table~\ref{TAB:CWEs}.
We found \NumberTotalCWEType{} different types of \gls{CWE}s in both platforms.

In ChatGPT we found \NumberGPTCWEType{} \gls{CWE}s.
The most frequent type of \gls{CWE}s are \textbf{CWE-327: Use of a Broken or Risky Cryptographic Algorithm} and \textbf{CWE-328: Use of Weak Hash}, both tied in first place with 95 snippets.
CWE-327 captures when cryptographic algorithms used are insecure. Hence, the desired security cannot be guaranteed.
Meanwhile, CWE-328 is similar as it covers when the hash algorithm does not meet security expectations.
The \gls{CWE} is found by the CodeQL rules \textit{``Use of a broken or risky cryptographic algorithm''} and \textit{``Use of a potentially broken or risky cryptographic algorithm''}.
Tied in second place is \textbf{CWE-404: Improper Resource Shutdown or Release} and \textbf{CWE-772: Missing Release of Resource after Effective Lifetime}, found in 37 snippets.
Both \gls{CWE}s cover when resources are incorrectly released.
CWE-404 captures a release before it becomes available, whereas CWE-772 is after the resource is no longer needed.
The \gls{CWE} is found by the CodeQL rules \textit{``Improper Resource Shutdown or Release''} and \textit{``Missing Release of Resource after Effective Lifetime'}.
Lastly, the third-most present is \textbf{CWE-798: Use of Hard-coded Credentials}. 
The vulnerability describes when credentials like passwords or cryptographic keys are hard-coded in the code and can be leveraged by attackers to bypass authentication.
Additionally, CWE-798 is in 18th place in MITRE's Top 25.
The rule \textit{``Hard-coded credential in API call''} captures the \gls{CWE}.

In \gls{SO} we found \NumberSOCWEType{} types of \gls{CWE}s.
Hence, ChatGPT generated fewer types of \gls{CWE}s with a difference of \NumberDifferenceTypeBoth{}.
Tied for first place, \textbf{CWE-327: Use of a Broken or Risky Cryptographic Algorithm} and \textbf{CWE-328: Use of Weak Hash} were found in 122 snippets.
In the second place, with 29 snippets, was \textbf{CWE-404: Improper Resource Shutdown or Release} and \textbf{CWE-772: Missing Release of Resource after Effective Lifetime}.
In third place was \textbf{CWE-798: Use of Hard-coded Credentials} with 20 snippets.
Interestingly, the order of the top most common \gls{CWE}s is the same for both ChatGPT and \gls{SO}.
However, the number of vulnerabilities for the most frequent \gls{CWE}s found in each platform differed with a delta ranging from 8 to 27.
For the top \gls{CWE}s, ChatGPT created fewer vulnerabilities for CWE-327 and CWE-328. 
Meanwhile, \gls{SO} produced fewer vulnerabilities for CWE-404, CWE-772, and CWE-798.

ChatGPT snippets were better than \gls{SO} for \NumberGPTCWETypeBetter{} types of \gls{CWE}s.
Meanwhile, \gls{SO} was better for \NumberSOCWETypeBetter{} types of \gls{CWE}s.
Finally, \NumberGPTCWETypeNeither{} types of \gls{CWE}s had the same number of snippets.
The \gls{CWE}s with the largest differences were  \textbf{CWE-327: Use of a Broken or Risky Cryptographic Algorithm}, \textbf{CWE-328: Use of Weak Hash}, \textbf{CWE-335: Incorrect Usage of Seeds in Pseudo-Random Number Generator (PRNG)}, and \textbf{CWE-798: Use of Hard-coded Credentials}.
CWE-335 occurs when a pseudo-random number generator is used, but the seed is incorrectly managed.
The vulnerability is captured by the rule \textit{``Random used only once''}.
The difference between CWE-327 and CWE-328 \gls{CWE}s was 27, while for CWE-335 and CWE-798 it was 16.
The remainder of \gls{CWE}s had differences ranging from 0 to 8 occurrences in the code snippets.

We also compared the statistical significance of each type of \gls{CWE} between the platforms using a Chi-squared test. 
We found five types of \gls{CWE} with statistical significance.
Ordered by the \gls{CWE}s with the most statistically significant results, 
\textbf{CWE-335: Incorrect Usage of Seeds in Pseudo-Random Number Generator (PRNG)} had the highest significance ($p <0.001$). 
ChatGPT generated the vulnerability 0 times compared to the 16 occurrences in snippets in \gls{SO}.
\textbf{CWE-835: Loop with Unreachable Exit Condition (`Infinite Loop')} had the second highest statistical significance ($p=0.005$). 
The vulnerability occurs when there is an infinite loop in an iterator.
The rule \textit{``Constant loop condition''} can cover the vulnerability.
The \gls{CWE} occurred in ChatGPT code 8 times, while 0 times for \gls{SO}.
\textbf{CWE-798: Use of Hard-coded Credentials} was the third highest statically significant difference ($p=0.03$). 
We found 36 occurrences in ChatGPT code, compared to the 20 occurrences in \gls{SO} code.
Lastly,
\textbf{CWE-570: Expression is Always False}
and 
\textbf{CWE-571: Expression is Always True} have statistically significant results with the same significance ($p=0.046$). 
The \gls{CWE}s capture when the evaluation of an expression always returns true or false, respectively.
The rule \textit{``Useless comparison test''} captures both \gls{CWE}s.
ChatGPT generated 0 \gls{CWE} instances compared to \gls{SO} that produced 4 for each.

We found three different \gls{CWE}s within MITRE'S Top 25 \gls{CWE}s for 2023.
Ranked by position, \textbf{CWE-078: Improper Neutralization of Special Elements used in an OS Command (`OS Command Injection')} is the highest ranked \gls{CWE} we found in 5th place.
The \gls{CWE} covers when an input is used in operative systems commands that an attacker could influence externally.
The vulnerability was captured by the rule \textit{``Executing a command with a relative path''}. 
Only 2 snippets generated by ChatGPT contained the \gls{CWE}.
Following is \textbf{CWE-476: NULL Pointer Dereference} located in the 12th place.
The vulnerability captures when a pointer deference occurs by expecting a valid pointer, yet the pointer is null.
The \gls{CWE} was present in 1 ChatGPT snippet and 6 \gls{SO} snippets.
Finally, \textbf{CWE-798: Use of Hard-coded Credentials} is the last \gls{CWE} we found on the list, positioned in 18th place.
We found the \gls{CWE} in 36 times in ChatGPT snippets, while 20 times in \gls{SO} snippets.
\section{Discussion}\label{SEC:Discussion}

\subsection{Developer recommendations}

As with any disruption to software development, there are potential benefits and risks with using \gls{LLM}.
Still, developers are under-educated on insecure code propagation from both platforms, as both contain vulnerable code that can propagate to developers.
We found \NumberUniqueVulOverall{} different vulnerabilities and \NumberTotalCWEType{} different types of \gls{CWE} in the platforms.
Adding another element to the software supply chain carries an inherent risk.
Information gathered by developers from an outside source should be handled with caution as it can become an attack vector for malicious actors.
Hence, our first recommendation is \textit{do not blindly trust code, \gls{AI}-generated or human-created, from outside sources}.
\gls{LLM} have provided an excellent opportunity to become more conscientious of the complete software supply chain.

Consequently, developers may wonder if they should even use \gls{LLM} and other web-based information sources due to security risks.
Our second recommendation is despite risks \textit{to use the platforms, but apply good software security practices}.
Static analysis tools and software testing can help detect copied and pasted code vulnerabilities.
Practices can be adopted starting with those with the highest impact~\cite{zahan2023Software}.
At the same time, developers need to pay more attention to CWE-335, CWE-570, and CWE-571 for LLM-generated code, while CWE-798 and CWE-835 for \gls{SO} answers.

\subsection{Future work}

There are several avenues for future work.
First, reducing insecure code can propagate fewer security risks to users.
In line with prior work~\cite{hong2021Dicos, fischer2022Nudging}, \textit{approaches to stop insecure code propagation} need further research.
Second, we found that ChatGPT generated less vulnerable code compared to \gls{SO} using CodeQL.
A question remains: \textit{why are there even any differences} as \gls{LLM} are trained using the internet, including sites such as \gls{SO}?
We hypothesize that the compression and aggregation of information of \gls{LLM} may reduce vulnerabilities.
Still, work must further study why such differences occur, validating or refuting our hypothesis.
Companies can leverage this finding by incorporating \gls{LLM} within their development tools or information sources, as there may be a possible added security benefit of the technology.
Finally, in our work, we focused on code generation.
Still, \gls{LLM} are incorporated in various software tasks, including code summarization and translation~\cite{zheng2023Towards}.
As such, future work should evaluate the \textit{security implications of \gls{LLM} for other software tasks}.
\section{Limitations}\label{SEC:Limitations}

The main limitation of our work is the representativeness and generalizability of our analysis with the diversity of questions and answers in software development. 
Our findings are limited by our sampling strategy, the number of code snippets analyzed, and the Java programming language of the snippets.
Future work can expand upon the generalizability of our findings by analyzing different development information sources, \gls{LLM}, and programming languages with more code snippets.
Another limitation is based on how we leveraged ChatGPT as our prompt may have biased our results, despite our mitigation efforts to construct and refine the prompt iteratively. 
Prompt engineering approaches can be investigated to potentially reduce vulnerabilities in the generated code.
At the same time, our findings are constrained by the time frame and version of ChatGPT analyzed, given the evolving nature of \gls{LLM} and their usage. 
Additionally, how we measured and detected vulnerabilities in the code snippets scope our findings.
Software vulnerabilities are a commonly used measure~\cite{morrison2018Mapping}, while static analysis tools to detect vulnerabilities are a common practice within software development~\cite{elder2022really}.
CodeQL, the tool we use to detect vulnerabilities, has been used in prior research to evaluate code generated by \gls{LLM}~\cite{pearce2022Asleep, fu2023Security}.
Static analyzers still generate false positives, though CodeQL is one of the less sensitive tools in Java~\cite{li2023Comparison}.
Lastly, the non-deterministic nature of \gls{LLM} limits the reproducibility of our work. 
To combat this threat, we make our dataset public.
\section{Related work}\label{SEC:Related}

Software supply chain attacks leverage software components to compromise downstream users~\cite{ladisa2023sok}. 
For example, relying on third-party packages from ecosystems like npm and PyPi could compromise consumer packages~\cite{ohm2020Backstabber, zahan2022Weak}. 
As the software supply chain continues evolving, a recent concern for practitioners is the usage of \gls{LLM} in development~\cite{enck2023S3C2,dunlap2023S3C2}.
Research has revealed that code generated by GitHub Copilot can introduce vulnerabilities in public repositories~\cite{fu2023Security}.
\gls{LLM} have thus become an additional consideration in the security of the software supply chain.

Regarding software security, works have leveraged \gls{LLM} for vulnerability repair~\cite{pearce2023Examining, huang2023Empirical} and detection~\cite{purba2023Software}.
Works have also shown uses for \gls{LLM} in software security testing~\cite{happe2023Getting, zhang2023How} and security code reviews~\cite{yu2024Security}.
Complementary to prior works, data poisoning~\cite{cotroneo2023Vulnerabilities} and prompt injection attacks~\cite{wu2023Deceptprompt} exploits have been found for \gls{LLM} in software development contexts.
Research has evaluated several quality attributes of code generated by \gls{LLM}, including the correctness and usability~\cite{nguyen2022Empirical, almadi2022How, vaithilingam2022Expectation, moradidakhel2023GitHub, liu2023Need, kabir2023Answers, xu2023We}.
At the same time, studies have evaluated the security of the generated code.
Insecure code propagation has previously been studied within software security for platforms as \gls{SO}~\cite{bai2019Qualitative}.
Research has identified security risks such as vulnerabilities~\cite{pearce2022Asleep, liu2023Need} and code smells~\cite{siddiq2022Empirical} in code generated by \gls{LLM}.
Datasets~\cite{siddiq2022SecurityEval} and frameworks~\cite{siddiq2023Generate} have also been created to evaluate the security of \gls{LLM} for code generation.

Prior work comparing the security of LLM-generated code is the most related to our study.
Research has contrasted differences in the code when developers use \gls{LLM}~\cite{perry2023Users, sandoval2023Lost}.
At the same, work has evaluated if \gls{LLM} have introduced the same vulnerabilities as humans~\cite{asare2023Copilot}.
In line with Asare et al.~\cite{asare2023Copilot} and Sandoval et al.~\cite{sandoval2023Lost}, the security impact of \gls{LLM} is low.
At the same time, Asare et al.~\cite{asare2023Copilot} also found that the vulnerabilities generated by \gls{LLM} were different than human-produced.
Contrary to the findings of the other works and our study, Perry et al.~\cite{perry2023Users} found that participants were more likely to generate less secure code.
Still, Perry et al.~\cite{perry2023Users} found participants who trusted \gls{AI} less provided code with fewer security vulnerabilities.
Hence, \gls{LLM} may be a positive addition to the security of software projects if leveraged intentionally.
The studies analyzed C, C++, Python, and JavaScript code, while we studied Java code.
Hence, we hypothesize that the findings are applicable across different programming languages.

Our study complements prior work by contrasting code vulnerabilities of \gls{LLM} with a web-based information source, \gls{SO}.
By comparing \gls{LLM} with an established platform like \gls{SO}, we enhance our understanding of the security risks associated with using generative \gls{AI}.
Hence, we help developers 
by increasing the awareness of the security risks when selecting the information source for code snippets.

\section{Conclusions}\label{SEC:Conclusions}

With the widespread adoption of \gls{LLM} in software engineering, developers have raised concerns about the security risk implications and the potential impact on the software supply chain. 
Developers are weighing the benefits and risks of using \gls{LLM} compared to other web-based information sources, such as online forums like \gls{SO}.
Notably, \gls{SO} also contains code with security issues. 
Hence, developers require empirical data comparing the security of both platforms to inform their choices.

We compared ChatGPT and \gls{SO} vulnerabilities for \NumberFinalSnippets{} Java code snippets using CodeQL.
Based on our findings, software developers are under-educated on insecure code propagation from any information online source, be it \gls{AI}-generated or human-created code.
ChatGPT as a platform generated more vulnerabilities and types of \gls{CWE} than \gls{SO}.
Still, the code in ChatGPT and \gls{SO} had \NumberUniqueVulOverall{} different vulnerabilities and overlapped only in \NumberOverlapPerVulnerabilitiesUnique{} of snippets.
\textit{Any code that can be copied and pasted from an outside source, \gls{AI}-generated or human-created, cannot be blindly trusted, requiring applying good software security practices.}
The security concerns surrounding generative \gls{AI} are an opportunity to increase conscientiousness about software security.

\section*{Acknowledgments}
This work was supported and funded by the National Science Foundation Grant No. 2207008, CNS-2026928, North Carolina State University Provost Doctoral Fellowship, and Goodnight Doctoral Fellowship.
Any opinions expressed in this material are those of the authors and do not necessarily reflect the views of any of the funding organizations.
We thank the Realsearch and WSPR research groups from North Carolina State University for their support and feedback.
Additionally, we are grateful for the reviewer's time and feedback. 

\balance
\bibliographystyle{ieeetr}
\bibliography{references}

\end{document}